\documentclass[aps,prl,reprint,superscriptaddress,showpacs]{revtex4-1}

\pdfoutput=1 
\usepackage{graphicx}
\usepackage{color}
\usepackage{booktabs}
\usepackage{bm}
\usepackage{longtable}
\usepackage{amsmath}
\usepackage{dcolumn}
\usepackage{placeins}
\usepackage{expdlist}

\begin{document}

\title {Linear and non-linear coherent coupling in a Bell-Bloom magnetometer}

\author{R. Gartman} 
\affiliation{National Physical Laboratory, Hampton Road, Teddington, TW11 0LW, United Kingdom}
\author{V. Guarrera} 
\email{v.guarrera@bham.ac.uk}
\affiliation{National Physical Laboratory, Hampton Road, Teddington, TW11 0LW, United Kingdom}
\affiliation{Midlands Ultracold Atom Research Centre, School of Physics and Astronomy, University of Birmingham,
Edgbaston, Birmingham B15 2TT, United Kingdom}
\author{G. Bevilacqua} 
\affiliation{Universit\`a di Siena, Siena, Italy}
\author{W. Chalupczak}
\affiliation{National Physical Laboratory, Hampton Road, Teddington, TW11 0LW, United Kingdom}

\date{\today}

\begin{abstract}

Spin-exchange collisions in hot vapours are generally regarded as a decoherence mechanism. In contrast, we show that linear and non-linear spin-exchange coupling can lead to the generation of atomic coherence in a Bell-Bloom magnetometer. In particular, we theoretically and experimentally demonstrate that non-linear spin exchange coupling, acting in an analogous way to a wave-mixing mechanism, can create new modes of coherent excitation which inherit the magnetic properties of the natural Larmor coherence. The generated coherences further couple via linear spin-exchange interaction, leading to an increase of the natural coherence lifetime of the system. Notably, the measurements are performed in a low-density caesium vapour and for non-zero magnetic field, outside the standard conditions for collisional coherence transfer. The strategies discussed are important for the development of spin-exchange coupling into a resource for an improved measurement platform based on room-temperature alkali-metal vapours.     
 
 \end{abstract}

\maketitle

\textit{Introduction.--} Generating and maintaining coherence is essential for many precision measurement techniques based on atomic spin manipulation. Spin-exchange collisions (SEC) can be employed to transfer coherence between atoms of different species or in different states \cite{Ruff1965}-\cite{Katz2015}. Coherence transfer has been demonstrated at high vapor densities and close-to-zero magnetic fields, in the so-called spin-exchange relaxation-free (SERF) regime, and at non-zero magnetic fields for atomic species with the same gyromagnetic ratios or following radiofrequency dressing of the atomic states. These schemes have been successfully implemented in a variety of different applications, including tests of fundamental physics \cite{Smiciklas2011} and cosmology \cite{Pospelov2013}, for quantum enhanced metrology \cite{Wasilewski2010}, medical diagnostics \cite{Xia2006}, and navigation \cite{Kornack2005}. Recently, transfer of higher order coherences has also been investigated and birefringence coherence has been shown to originate from the Larmor coherence within the same hyperfine ground state by a non-linear spin-exchange collisional process \cite{Katz2013}. Coherences of higher order than the Larmor are interesting as they are involved in phenomena such as the non-linear magneto-optical rotation, which have been used for improved magnetometry schemes \cite{Pustelny2008}. 
\begin{figure}
\centering
\includegraphics[width=0.48\textwidth]{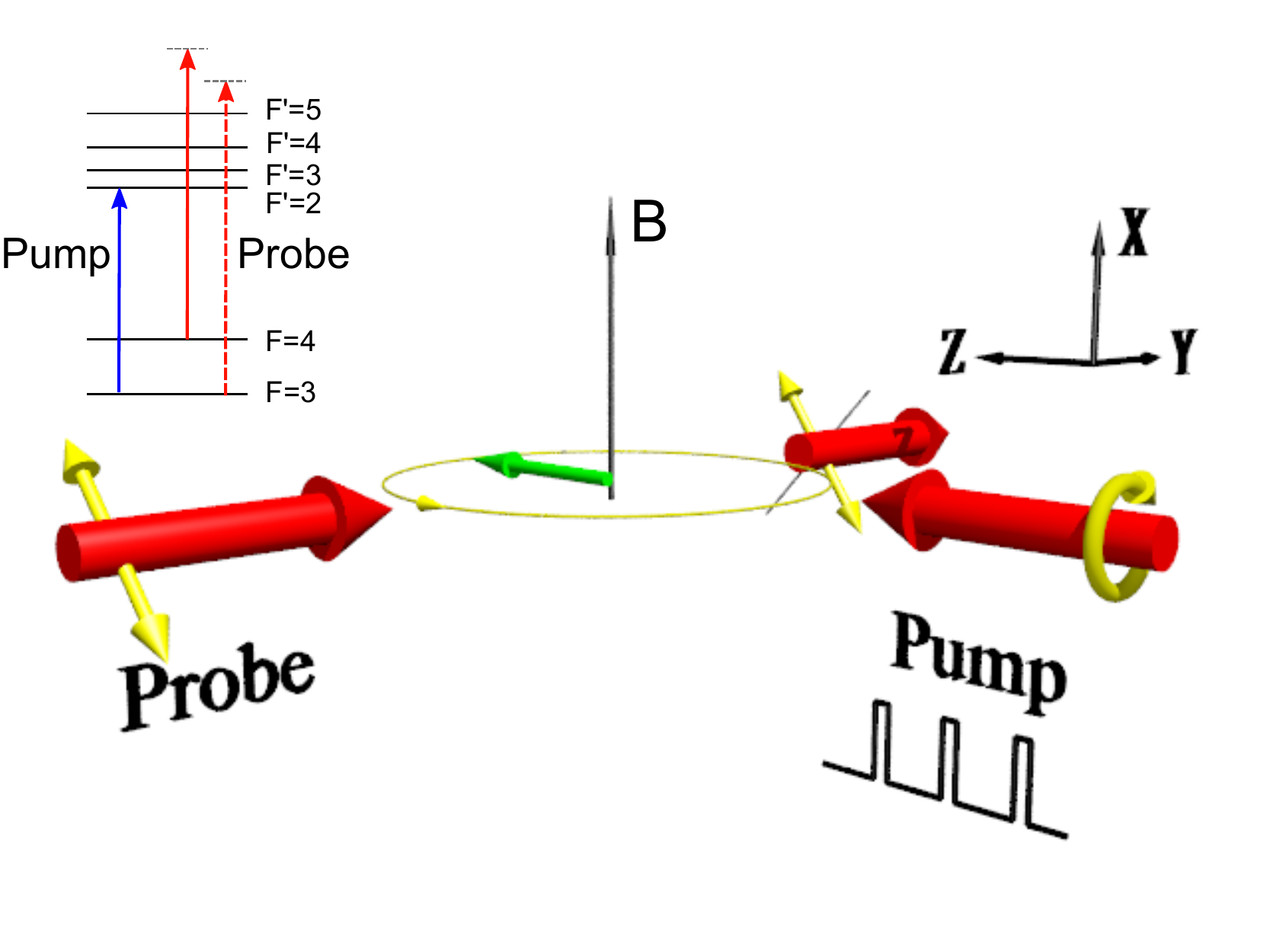}
\caption{ (color online) Schematics of the experiment: a circularly polarized pump with periodically modulated amplitude creates a collective atomic spin (green arrow) orthogonal to the applied DC magnetic field (thin black arrow). The atomic spin precession is then probed by detecting the polarization rotation undergone by a linearly polarized off-resonant probe, propagating through the atomic sample orthogonally with respect to the pump beam and the applied magnetic field. The relevant transitions for pumping/probing on the $^{133}$Cs D$_2$ line are shown. }
\label{fig:Setup}
\end{figure}

In this work, we experimentally demonstrate that the coherence oscillating at the Larmor frequency $\omega_{\text{L}}$ (natural coherence) can be effectively transferred to new \textit{controllable} modes (secondary coherences). Magnetic properties of the former are transferred to the latter across different hyperfine states, notably in the regime of low atomic density and non-zero magnetic field. These conditions fall outside the standard regime for coherence transfer based on linear SEC coupling. We explain this effect as being due to the combination of the non-linear coupling of the SEC term and the amplitude modulation of a relatively strong pump in a Bell-Bloom scheme ($\Omega_R \gg \omega_L $, with $\Omega_R$ being the Rabi frequency). We finally observe that the natural and secondary coherences can further interact in a linear fashion leading to $15 \%$-$25 \%$ narrowing of the Larmor coherence widths. This behaviour reveals that modulation strategies might represent an important resource for manipulating coherences in optically-pumped magnetometers, and that non-trivial resonance structures can appear even at low atomic densities when the pumping term is sufficiently strong. 
\begin{figure*}
\centering
\includegraphics[width=0.48\textwidth]{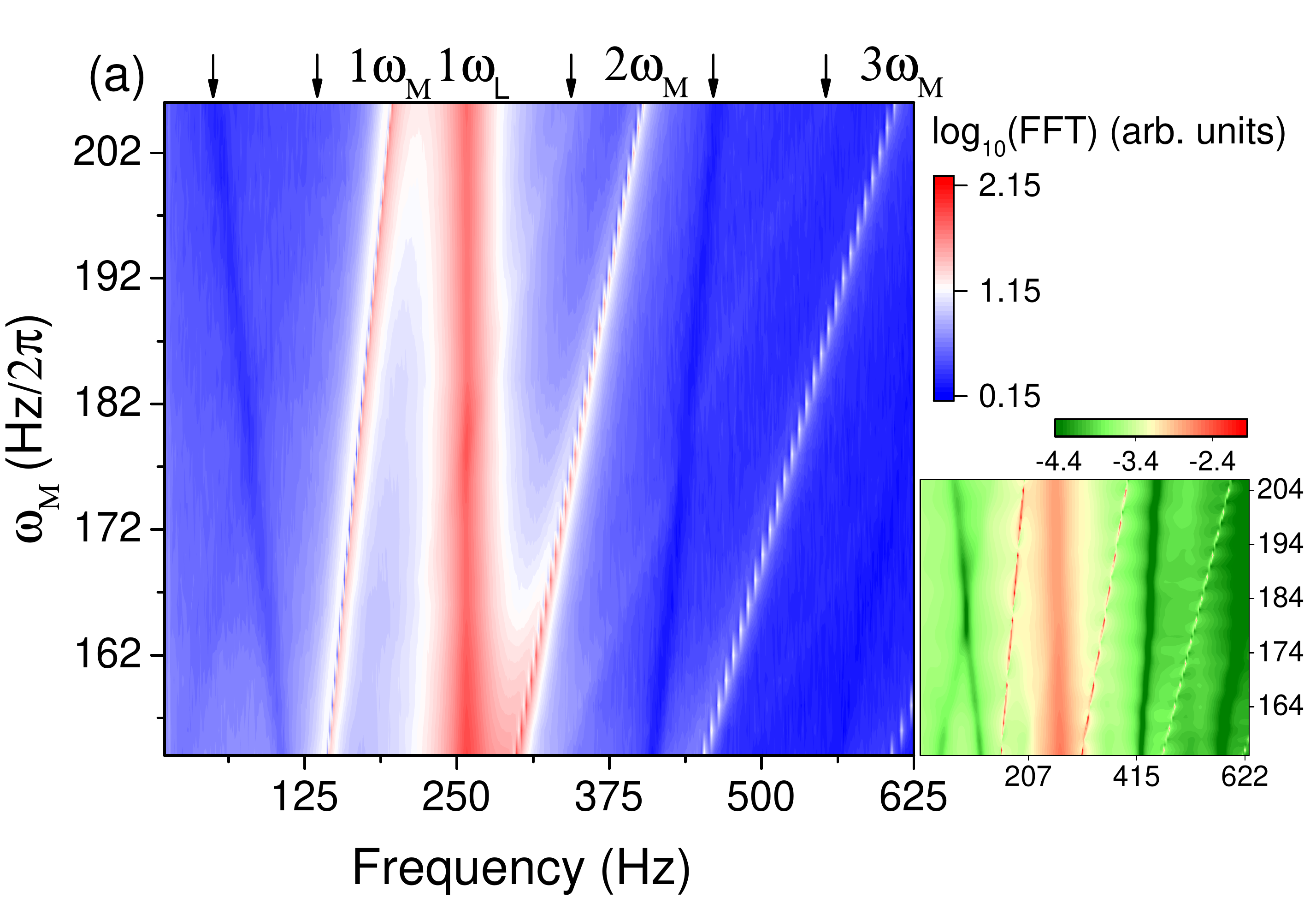}
\includegraphics[width=0.48\textwidth]{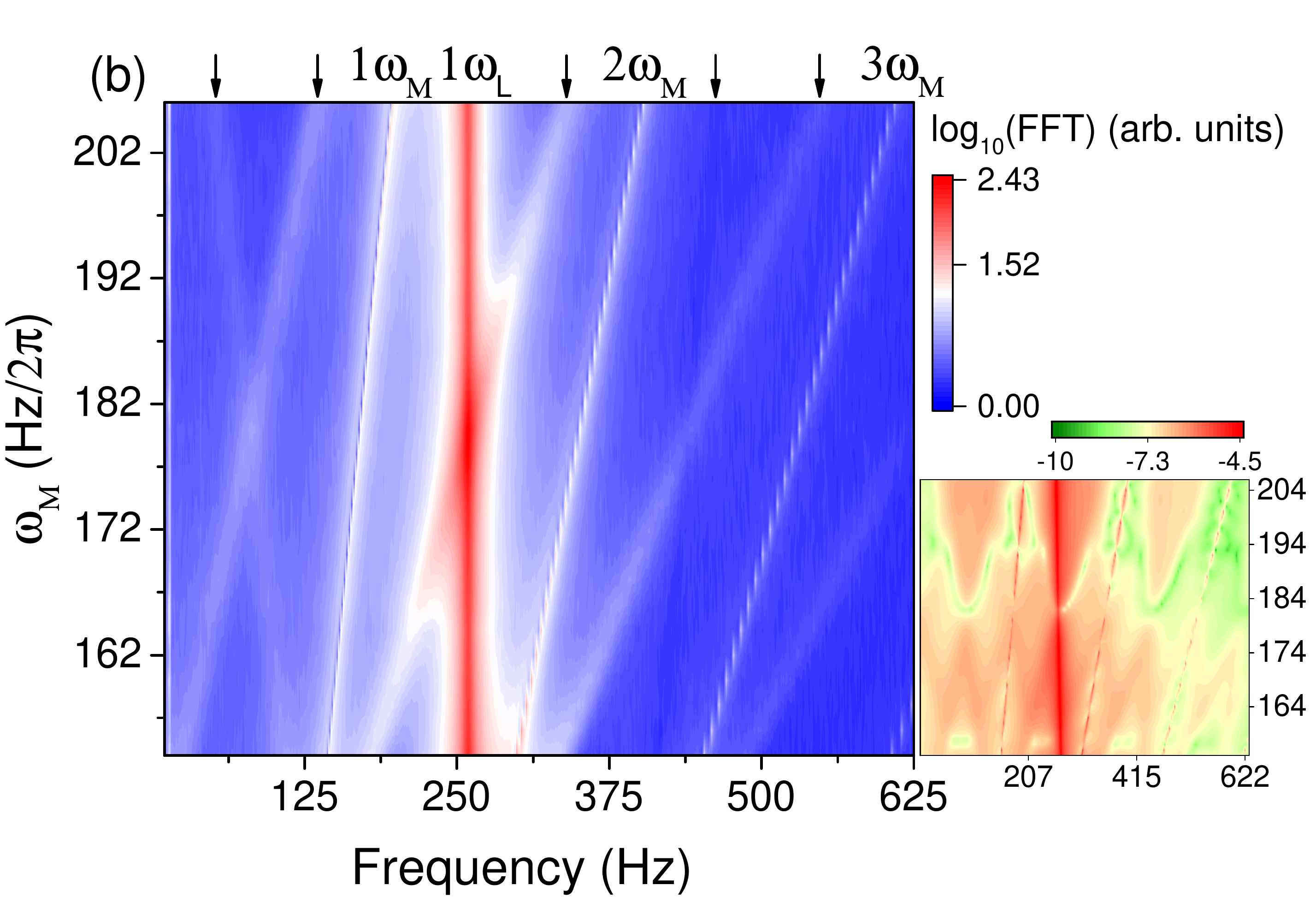}
\caption{ (color online)  FFT spectrum of the magneto-optical rotation signal detected for atoms in (a) $F=3$ and (b) $F=4$ for different frequency of modulation of the pump beam power. This portion of the spectrum shows the evolution around the Larmor frequency $\omega_L/2 \pi \sim 270$ Hz. Together with the peak at $\omega_L$ and the first three harmonics of $\omega_M$, additional peaks (and dips) depending on $\omega_M$ are visible around integer multiples of $\omega_M/2$ (highlighted by black arrows in the figure). The pump and probe beam powers have been set to 80 $\mu$W and 600 $\mu$W  and for each modulation frequency the spectrum is the result of over 100 different acquisitions. Insets: calculated FFT spectrum of the average macroscopic spin component along the direction of the probe beam ($\widehat{y}$ direction) for atoms in $F=3$ and $F=4$. The pump power has been set to $\Omega_R=100 \omega_M$, which corresponds to $ 20 \mu$W power for our experimental settings. Due to computational issues we cannot simulate higher pump powers, for which we have compensated by using a stretched state to initiate the $F=3$ ground state, mimicking saturation by a strong pump.}
\label{fig:Spectrum}
\end{figure*}

\textit{Experimental system.--} We study the magneto-optical signal generated by a collective spin of Cs atoms in the $F=3,4$ ground states, precessing around a DC magnetic field. Measurements are performed in a low density ($0.33 \times10^{11} \text{cm}^{-3}$) thermal vapour housed in an anti-relaxation, paraffin coated, glass cell. The atomic coherences are generated by a Bell-Bloom pumping process \cite{Bell1961, Gartman2015}. In this scheme a train of optical excitation pulses produces atomic coherences within both the ground state manifolds. The $F=3$ level is driven by resonant optical pumping and the $F=4$ level is pumped mainly via SEC and optical off-resonance excitation \cite{Gartman2015}. We focus our analysis on the initial phase of the pumping process. As pointed out in \cite{Grujic2013} the spectrum of the evolving ground state coherences contains, in that case, two components: (1) a steady-state oscillation at the modulation frequency $\omega_M$, and (2) a transient oscillating at $\omega_{\text{L}}$ damped on a timescale related to the SEC rate. In order to address them independently, the modulation of the pump is performed outside the standard frequency range of interest of a Bell-Bloom scheme, at a frequency $\omega_M$ whose integer multiples do not overlap with $\omega_L$, i.e. $n \times \omega_M \neq \omega_{\text{L}}$. In our setup, the atomic coherences are monitored by continuous Faraday-type polarization rotation measurements \cite{Takahashi1999}. The polarization of the probe beam transmitted through the vapour cell is analysed by a polarimeter. The magneto-optical rotation signal collected by a balanced photodetector is then analysed to extract its Fourier transform, which, along with the continuous measurement, enables monitoring of the spectral components oscillating at frequencies other than the modulation $\omega_M$. 
Details of our experimental setup can be found elsewhere \cite{Gartman2015}, we just recall here that the pump is obtained by a circularly polarised laser beam, frequency locked to the caesium $6\,^2$S$_{1/2}$ $F=3 \rightarrow{}6\,^2$P$_{3/2} F'=2$ transition (D$_2$ line, 852 nm). The pump power is modulated with a square pulsed waveform and duty cycle of $7\, \%$. The signal produced by the $F=4$ ($F=3$) ground state atomic coherences is read out by a probe beam propagating in a direction orthogonal to the pump beam, frequency locked to the $6\,^2$S$_{1/2}$ $F = 4 \rightarrow{}6\,^2$P$_{3/2} F' = 5$ ($6\,^2$S$_{1/2}$ $F =3 \rightarrow{}6\,^2$P$_{1/2} F' = 4$) transition, and subsequently frequency shifted by $960$ MHz to the blue side by two acousto-optic modulators in a double-pass configuration. The measurements are performed by scanning the frequency of the pump modulation across the value $\omega_0 = \frac{2}{3} \times \omega_{\text{L}} \simeq 2\pi \times 180$ Hz, which corresponds to the third harmonic of the square-shaped modulation coinciding with $2 \: \omega_L$ \cite{modulation}. 

\textit{Non-linear coupling.--} The Fourier transform of the magneto-optical signal is shown in Fig.~\ref{fig:Spectrum}(a)-(b) for the $F=3$ and $F=4$ states respectively. The peaks at $\omega_L \simeq 2\pi \times 270$ Hz, corresponding to the transient oscillation are clearly visible, together with the first three harmonics of the square-shaped modulation function. Additionally, we observe several secondary peaks (and dips), appearing around the semi-integer multiples of the modulation frequency, whose positions shift as a function of $\omega_M$ and merge when $\omega_M = \omega_0$ according to the scaling $\omega_{\pm}= \frac{n}{2} \omega_M \pm \frac{1}{2}(3 \omega_M-2\omega_L)$, with $n$ an odd integer. The signal shown in Fig.~\ref{fig:Spectrum} (a)-(b) was recorded during the first $4$ seconds from the beginning of the optical pumping, however we have verified that the contribution to the observed features comes solely from the transient dynamics, i.e. roughly the first $300$ ms from the start of the pumping process when the induced natural coherences are present \cite{Footnote1}. 

\begin{figure*}
\includegraphics[width=0.32\textwidth]{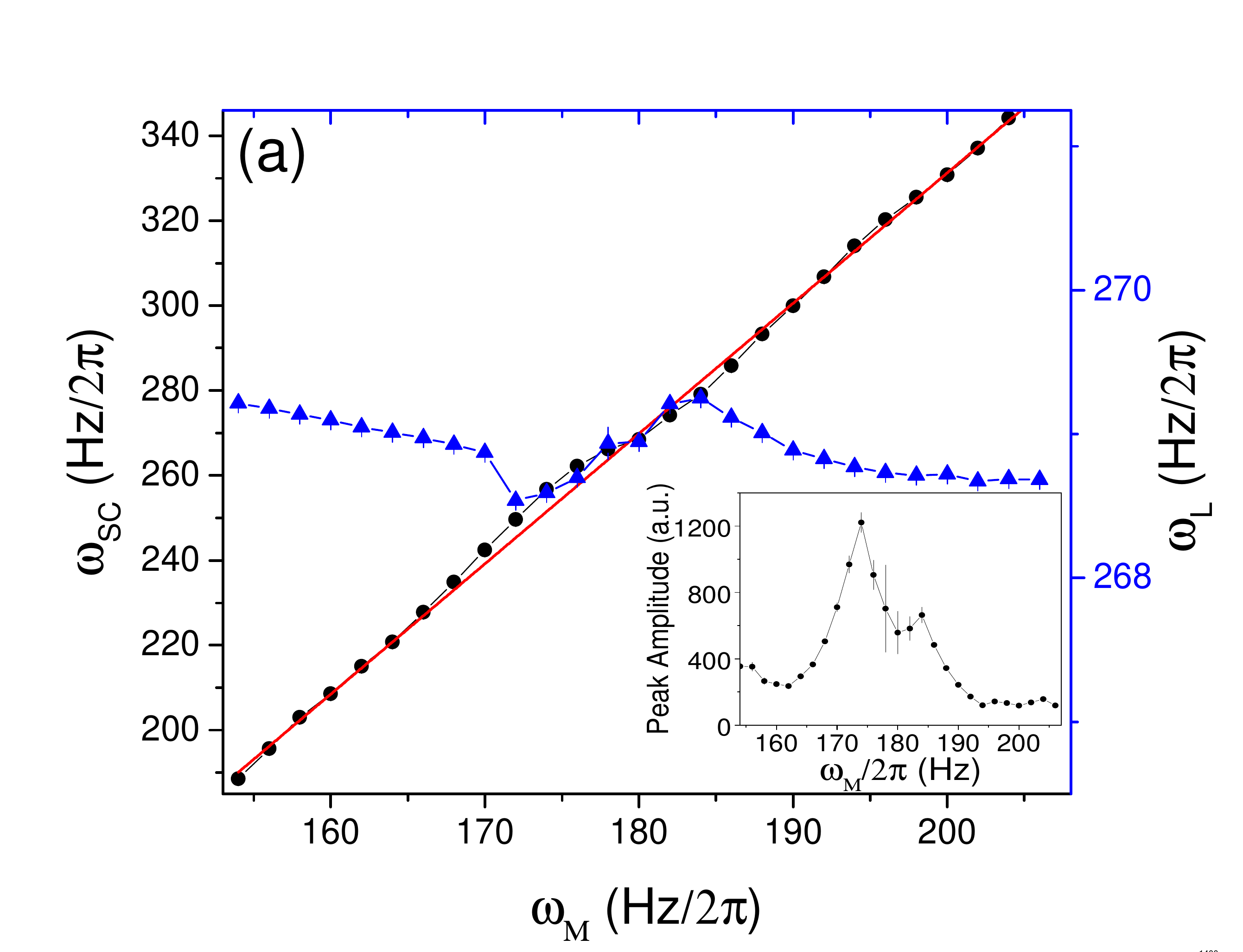}
\includegraphics[width=0.32\textwidth]{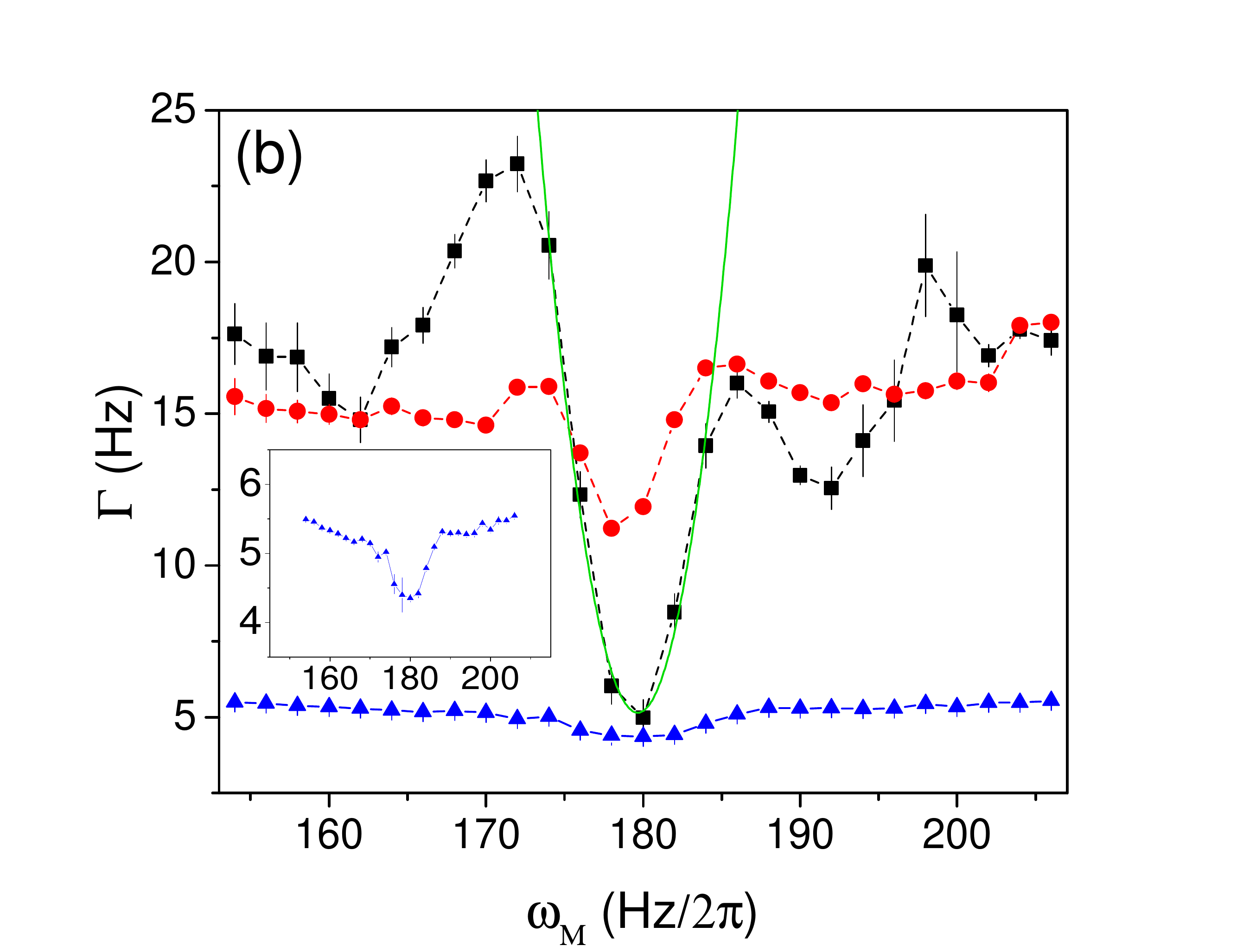}
\includegraphics[width=0.32\textwidth]{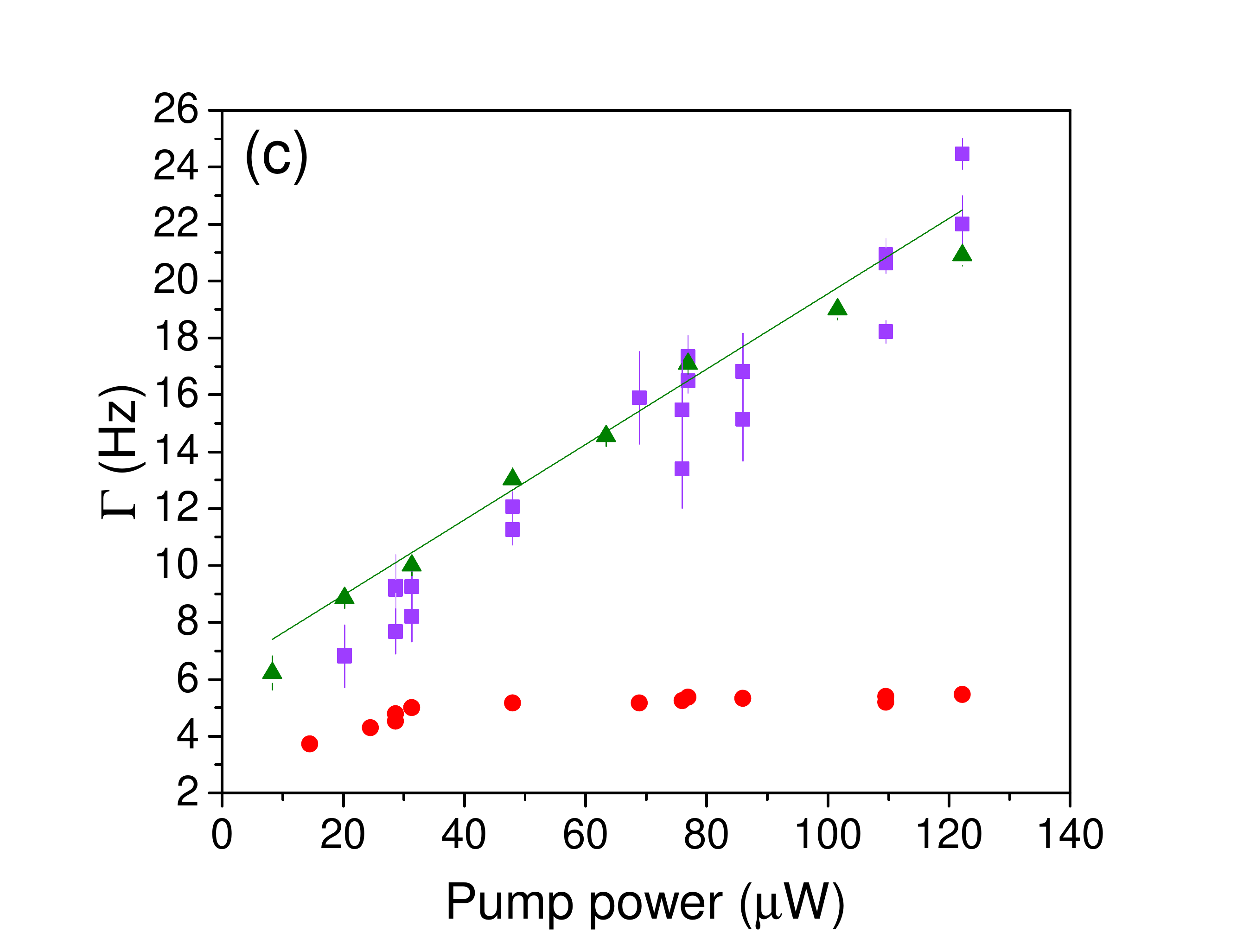}
\caption{ (color online) (a) Positions of the FFT secondary coherence profile crossing $\omega_L$ (see text) as a function of the pump modulation frequency (black dots, scale on the left) and position of the natural coherence profile (blue triangles, scale on the right) for atoms in F=4. The red solid line is a linear fit to the secondary coherence data providing the functional dependence $\omega_{SC}=(3.06 \pm 0.02) \omega_M -(272 \pm 2)$ Hz $\sim (3 \omega_M -\omega_L)$. The inset shows the fitted amplitude of the secondary coherence profile as a function of the modulation frequency. (b) Linewidth of the natural coherence for atoms in F=4 (blue triangles), secondary coherence for atoms in F=4 (black squares), and Larmor coherence for atoms in F=3 (red dots) as a function of the modulation frequency. The green solid line shows a quadratic fit to the secondary coherence data with function $\Gamma=(0.16\pm 0.01)(3 \omega_M-2\omega_L)^2 + (5.1 \pm 0.4)$ Hz$ \sim (\omega_{SC}-\omega_L)^2 + \Gamma_0$, where we have used the linear fit above for substituting $\omega_{M}$. The inset of the figure is a zoom into the natural coherence linewidth for F=4, showing a reduction of the peak size at resonance $\omega_M=\omega_0=180$ Hz. (c) Linewidth as a function of the pump beam power for the natural coherence of F=4 atoms (red dots), the secondary coherence for F=4 atoms (purple squares), and the natural coherence for F=3 atoms (green triangles). For the secondary coherence profile in F=4, measurements are taken on one side of the resonance with $160 < \omega_M < 170$ Hz. The green solid line is a linear fit to the F=3 data. Each point in the above figures is obtained from the average of 100 acquisitions.}
\label{fig:Position}
\end{figure*}

To better characterize the nature of the observed peaks, in Fig.~\ref{fig:Position}(a)-(b) we report the position and linewidth of the secondary coherence profile moving accross $\omega_L$ in $F=4$ as a function of the modulation frequency \cite{Footnote2}. We observe that its position ($\omega_{SC}$) varies approximately linearly with the third harmonic of the optical excitation (fitted slope is $3.06 \pm 0.02$), in other words the rate of change of its location follows the harmonic of the pump modulation around the birefringence coherence at $2 \omega_L$. 
In addition, the linewidth $\Gamma$ away from the resonance at $\omega_M=\omega_0$ for the atoms in $F=4$ is compatible with the linewidth of the natural coherence profile measured for the atoms in $F=3$. We recall that, because the pumping scheme implements direct optical excitation in the F=3 and off-resonant transfer to the F=4 level, the linewidths of the atomic coherences in these two ground state levels differ significantly. This dependence is further confirmed by varying the power of the pump beam, see Fig.~\ref{fig:Position}(c). Higher power broadening of the $F=3$ coherence linewidth due to direct optical excitation can be seen in the linewidths of all the secondary peaks in $F=4$. Transfer of characteristic features, such as the linewidth, from the $F=3$ to the $F=4$ level reveals transfer of coherences between the two different hyperfine states.  
Coherences in the alkali-metals' ground state levels oscillate at opposite frequencies and there cannot be transfer due to linear SEC coupling between freely evolving modes. In the presence of a modulated pump, linear coherence transfer due to locking of the natural coherence to the external frequency drive has been demonstrated but only between different atomic species and the magnetic properties of one species, including its coherence time and gyromagnetic ratio, were not transferred to the other \cite{Skalla1996}. 

To provide intuitive insight into the nature of the SEC coherent coupling, we analytically solve the dynamics for a simplified system:
\begin{equation}
\frac{d \rho}{d t} = (W+Z+E) \rho + \epsilon Q(\rho) \rho +f(t)V
\label{eq:tre}
\end{equation}       
where the operators $W+Z+E=L$ denote the linear part of the Liouville equation \cite{Happer1977}, taking into account the hyperfine structure interaction $W \rho = A_{hfs}/(i\:\hbar) \left[ \textbf{I} \cdot \textbf{S}, \rho \right]$, the external magnetic field interaction $Z \rho =  \omega_L/(i\:\hbar) \left[  S_x, \rho \right]$, and the linear SEC interaction $E \rho=-R_{SE} \left( \textbf{A} \cdot \textbf{S} \right)$. In the expressions above $A_{hfs}$ is the strength of the hyperfine interaction, $S_x$ is the component of the electronic spin along the direction of the magnetic field $\textbf{B}=B\textbf{x}$, and $R_{SE}$ is the rate of SEC. The term due to the non-linear SEC is $Q(\rho) \rho = R_{SE} ( 4 \alpha \textbf{S}\cdot \langle \textbf{S} \rangle ) $, which is weighed by an arbitrarily small amplitude $\epsilon$ in this perturbative approach. The operators $\alpha$ and $\textbf{A}$ are the nuclear and electronic parts of the density matrix $\rho=\alpha+\textbf{A} \cdot \textbf{S}$. The simplest description for the amplitude modulated pump, allowing an analytical treatment, is obtained by introducing the additive term $f(t)V$ where $f(t)=\sum_n f_n e^{i n \omega_M t}$, $V$ is a diagonal matrix, and there is no direct dependence on $\rho$. We search solutions of Eq.\ref{eq:tre} of the form $\rho(t)=\rho_0(t)+\epsilon \rho_1(t)+...$ by solving the coupled equations:
\begin{align}
\rho_0&=L\rho_0+f(t)V  \nonumber \\
\rho_1&=L\rho_1+Q(\rho_0) \rho_1. 
\label{eq:quattro}
\end{align}  
To derive the explicit expressions for the operators the density matrix is expanded as $\rho = \sum_{LM, FF'} \rho_{LM}(FF') \vert LMFF'\rangle$, by using the coupled spherical basis operators $\vert LM FF'\rangle$ (see Eq.~60 in \cite{Happer1977}). After simplifying the notation so that $\rho_{LM}(FF')=\rho \textunderscore i$, one can find that $\rho_{0} \textunderscore i =A_i e^{\lambda_it} + \sum_n w_{n,i} e^{in\omega_Mt}$, while $\rho_{1} \textunderscore i$ is a sum of different modes with frequencies:
\begin{align}
\lambda_i \quad & \nonumber \\
\lambda_j+\lambda_k \quad & \text{for} \quad \lambda_j+\lambda_k-\lambda_i \neq 0 \nonumber \\
i\: (n+m) \omega_M \quad & \text{for} \quad -\lambda_i+ i\: (n+m) \omega_M \neq 0 \nonumber \\
\lambda_j+ i\: n \omega_M \quad & \text{for} \quad \lambda_j-\lambda_i+im\omega_M \neq 0 
\label{eq:cinque}
\end{align}  
where the eigenvalues $\lambda_i$ can be written as $\lambda_i=-\Gamma_i -i\: \omega_i$ \cite{Happer1977,Katz2013}. The last term in Eq.~\ref{eq:cinque} shows that the non-linear SEC term in the presence of the pump modulation leads to the generation of new modes with frequencies depending on $(\omega_j+n\omega_M)$ and with magnetic properties inherited from the natural coherences of the system $\Gamma_j$. In other words, non-linear SEC induce a mode mixing of the magnetic multiplets and the stationary modes driven by the pump at the modulation frequency. This mode mixing, which cannot be found when only considering the linear part of Eq.~\ref{eq:tre}, is compatible with what is observed in the experiment. 
To further confirm the results of our simple model, we have performed numerical simulations starting from the full master equation:
\begin{align}
 \frac{d \rho}{d t} & =  (W+Z+E) \rho + Q(\rho) \rho + \Lambda(t)
\label{eq:uno}
\end{align}
where the simple pumping term of Eq. \ref{eq:tre} has been replaced by
\begin{equation}
\Lambda(t)  =- L_R \left\lbrace WW^{\dagger}, \rho \right\rbrace  +2 L_R W \rho_{e} W^{\dagger} 
+ (\rho')_{sp}
\label{eq:due}
\end{equation}
where $W=-E_0(t) \Pi \textbf{d} \cdot \textbf{e}^* \Pi_e$ ($W^{\dagger}=E_0(t) \Pi_e \textbf{d} \cdot \textbf{e} \Pi$), with $E_0(t)$ and $\textbf{e}$ the amplitude and polarization versor of the laser field respectively, and $\textbf{d}$ the induced atomic dipole moment. $\Pi$ and $\Pi_{e}$ are the projectors on the ground and excited state (which is labelled with the letter $e$), and $L_R$ is a coefficient depending on the natural linewidth of the transition. This term describes resonant pumping of the $F=3$ level, while optical off-resonant pumping into the $F=4$ level is neglected \cite{nota}. The first term in Eq.~\ref{eq:due} accounts for depopulation of the ground state, while the second and third refer to repopulation due to stimulated and spontaneous emission. Using the expression obtained in \cite{Ducloy1970}, the latter term can be expanded into multiplets $\frac{d}{dt} \rho_{LM}(FF) = \sum_{F_e} \xi_L(F_e \rightarrow F) \rho_{LM}(F_eF_e)$ with: 
\begin{align}
\xi_L(F_e \rightarrow F) = & (-1)^{F_e+F+L+1} (2J_e+1)(2F_e+1)(2F+1) \nonumber \\ 
&\Gamma(J_e \rightarrow J) \left\lbrace 
\begin{matrix}
F_e & F & 1 \\
J & J_e & I
\end{matrix}
\right\rbrace ^2
\left\lbrace  
\begin{matrix}
F & F & L \\
F_e & F_e & 1
\end{matrix}
\right\rbrace 
\end{align}
where $\Gamma(J_e \rightarrow J)$ is the natural linewidth of the transition, and $\left\lbrace  \right\rbrace $ is the Wigner 6-j symbol. We note that in the calculations we neglect the hyperfine coherences in the ground state and we include only the $L=0,1,2$ terms in the expansion of the density matrix.
The results of the numerical simulations, shown in the insets of Fig.~\ref{fig:Spectrum}, show good agreement with the main features of our experimental data. Additionally, they confirm that: 1) the non-linear coupling term is essential for explaining the new features appearing in the spectrum, and that 2) the non-additive terms in the pump of Eq.~\ref{eq:due} are instrumental for recovering the precise  frequency scaling which is experimentally observed. Concerning the latter point, we note that we can detect these new features only for relatively strong pump power for which a simple description by an additive term is not appropriate anymore.

\textit{Linear coherence transfer.--} For linear SEC-driven coherence transfer it is essential that the coherences that are transferred in sequential random collisions are in phase. This imposes a resonance condition on the frequencies of the modes involved in the transfer, $\omega_{\text{A, B}}$, with respect to the time between the collisions $\tau_{\text{SEC}}$, i.e. $(\omega_{\text{A}}-\omega_{\text{B}}) \times \tau_{\text{SEC}} <1$  \cite{Haroche1970}. This condition is met when the spectral lines created by the atomic coherences, and coupled by SEC, overlap \cite{Chalupczak2014}. The range of detuning where coherence transfer can be observed is thus roughly defined by the SEC-limited width of the relevant spectral lines. As already discussed, the resonant condition is normally achieved in experiments by simultaneously adjusting $\omega_{\text{A, B}}$ with a magnetic field \cite{Happer1973, Happer1977, Chalupczak2014, Katz2015}. Characteristic signatures of coherence transfer are known to be (a) a reduction of the distance $(\omega_{\text{A}}-\omega_{\text{B}})$ between the modes generated by coherence oscillation and (b) narrowing of the relevant profile linewidths following a quadratic dependence on the detuning $\Gamma \sim (\omega_{\text{A}}-\omega_{\text{B}})^2$ \cite{Happer1977}. In our case the strength of the coupling is not controlled by the magnetic field but by the modulation frequency of the pump amplitude. 
We focus our analysis on the secondary coherence profile $(\omega_A=\omega_{SC})$ moving across the natural coherence $(\omega_B=\omega_L)$. In Fig.~\ref{fig:Position}(a)-(b), signatures of the linear SEC coherence transfer can be observed. Firstly, a detailed inspection of the change of the position of the secondary coherence peak (black dots) with $\omega_M$ reveals a deviation from the linear fit, shown with a solid red line in Fig.~\ref{fig:Position}, well appreciable out of the measurement noise. This is mirrored by a change in the position of the natural coherence profile (blue triangles) within a $\pm 0.5$ Hz range. Note that in the same interval of $\omega_M$ the secondary peak is also significantly amplified, as is clearly visible in Fig.~\ref{fig:Spectrum} and in the inset of Fig.~\ref{fig:Position}(a). Secondly, the secondary peak linewidth is significantly reduced and shows compatibility with a quadratic dependence on the detuning from the resonance ($\omega_{SC}-\omega_L$), as shown in Fig.~\ref{fig:Position}(b). On resonance, where the coupling is strong, the secondary coherence profile completely inherits the longer coherence time $T_2=1/\Gamma_B$ of the dominant natural coherence ($\Gamma_B \ll \Gamma_A $) which is mainly determined by SEC \cite{Katz2015}. Notably, narrowing of the natural coherence profile linewidth by $15\%$-$25 \%$ is also observed around the resonance for our experimental parameters. 
This suggests that the coupling of the natural and secondary coherence, which is a mixture of atomic multiplets of different hyperfine levels, counteracts the coherence relaxation due to SEC at the Larmor frequency, effectively increasing the associated coherence time $T_2$. This is similar to the linear SERF effect, as it follows a SEC-induced coherent hybridization of the quantum states of the atoms between the two hyperfine levels. However, unlike SERF we observe this effect in the unrestrictive regime of non-zero magnetic field and for rather low atomic density. Importantly this demonstrates that in the case $T_2$ is limited by SEC, the controlled generation of new hybrid coherences obtained in a \textit{non-resonant} way can be used to induce SERF-like coherence improvement of $\Gamma$ out of the typical SERF regime, and enhance the operation of atomic magnetometers.

\textit{Conclusions.--} We have demonstrated that the magneto-optical signal of a Bell-Bloom magnetometer shows a rich spectrum dependent  on the modulation frequency when a relatively strong pump is applied. This allows coherence transfer across different hyperfine states, a situation which cannot be achieved with linear SEC coupling in a regime of non-zero magnetic field and low atomic density, without further magnetic manipulation of the atomic states. We identify this effect as due to non-linear coupling, as resulting from non-linear SEC. The new coherences are thus created in a process that resembles a four-wave mixing scheme, led by collisional interactions between the atoms for our off-resonant pumping scheme - a situation previously observed only in momentum states of ultracold atoms \cite{Perrin2007}. The new peaks still linearly couple to the natural oscillation at $\omega_L$, leading to narrowing of the observed spectral profile and modification of its position. Our work shows that the combination of a controlled modulation and a non-linear coupling term provides a rich scenario that can be exploited for improving spin coherence, and paves the way to new schemes for generating spin-squeezing in room-temperature vapour magnetometry \cite{chapman2013}.

\begin{acknowledgements}
The work was funded by the UK Department for Business, Innovation and Skills as part of the National Measurement System Programme. V.G. was supported by the EPSRC-UKRI (EP/S000992/1). We thank G. Barontini, R. Godun, and R. Hendricks for critical reading of the manuscript. 
\end{acknowledgements}


\begin{thebibliography}{}
\bibitem{modulation} We note that the results obtained do not qualitatively depend on the chosen modulation frequency, provided the condition $n \times \omega_M \neq \omega_{\text{L}}$ is fulfilled.
\bibitem{Footnote1} Observation of the weak stationary component at $\omega_{\text{L}}$ will be discussed elsewhere.
\bibitem{Footnote2} Analogous components and similar behaviour are observed in the F=3 level.
\bibitem{nota} To clarify the role of off-resonant pumping to the $F=4$ state in generating the observed peaks, we have repeated the measurements with the pump frequency red-detuned with respect to the $F=3 \rightarrow F'=2$ transition by approximately $300$ MHz (Doppler broadening is roughly $230$ MHz). No qualitative difference was observed in the collected data, which justifies the predominance of SEC coupling between the $F=4$ and $F=3$ states.
\end{thebibliography}
\end{document}